\documentstyle[12pt]{article}
\def\d{{\rm d}}
\begin{document}
\begin{flushright}
OCHA-PP-64

March 1995
\end{flushright}
\centerline{Multiplicity and Event Shape in the Perturbative QCD}
\vspace{0.5cm}
\centerline{K. Tesima}
\centerline{\footnotesize Department of Physics, Ochanomizu University}
\centerline{\footnotesize Ohtsuka, Bunkyo, Tokyo 112, Japan}
\centerline{ Invited talk at the 3rd Workshop on TRISTAN}
\centerline{Physics at High Luminosities (October 1994)}
\vspace{0.3cm}

\begin{abstract}
The multiple hadroproduction in the perturbative QCD
is briefly reviewed.
There are a number of quantities which can be analysed
with the use of the high-luminosity TRISTAN data.
The analysis will contribute to clarifying some unsolved questions,
and to the deeper understanding of the jet physics.
\end{abstract}

\vspace{0.3cm}

When a coloured particle is produced at a high energy,
a number of hadrons are produced predominantly near the direction
of the high energy particle (jet phenomenon).
The multiplicity of hadrons in a jet increases at higher energies.
In a jet from an energetic quark at 500GeV, for example,
we expect to find sixty or more hadrons in average.
In the case of a gluon-jet,
the multiplicity will be about twice as many.
It is impossible to obtain the precise knowledge of the production
of the individual hadrons of this large number.

The multiple hadroproduction would give a large background
when we try to find new physics in high energy reactions.
It is highly desirable to have its quantitative understanding.
In place of evaluating the precise exclusive cross section,
in which all the individual hadrons are specified,
we have to investigate more inclusive quantities,
such as the multiplicity or the hadron distributions.

The modified leading-log approximation (MLLA) is the theoretical
framework which enables us to systematically analyse
the multiparticle production in the perturbative QCD.
In the first part of this talk, I'll give a very brief
description of MLLA and the related theoretical aspects.
Although it requires lengthy and complicated mathematical expressions
to give the full representation of the theory,
the physical picture behind it is very simple and clear.
More precise expressions and their detailed derivation are found
in the standard review \cite{dok1}
and in the original papers referred in this talk.

\section{The Theory of Multiple Hadroproduction}

In order to obtain quantitatively precise theoretical results
from the first principle,
we have to rely mostly on the perturbation theory at this moment.
The reason that we can make use of the perturbation theory for
the {\em strong} interaction (QCD) is indeed its asymptotic freedom.
The effective QCD coupling $\alpha_s(Q^2)$ is
\begin{equation}
\alpha_s(Q^2)\approx \frac{1}{b_0\ln (Q^2/\Lambda^2_{QCD})}\;,
\end{equation}
where $b_0=23/(12\pi)$ (with five active quark flavour),
and $Q$ is the typical momentum size of the interaction.
At the asymptotically high energy,
$\alpha_s(Q^2)$ provides the infinitesimal parameter
for the perturbation theory.

The straightforward perturbation theory, however, would fail
if we tried to apply it to the multiple hadroproduction.
Because QCD has the infrared (IR) singularity,
the majority of the particles emitted
from an energetic parton are soft gluons
(\lq \lq soft'' means much lower energy than the total energy of the
hard interaction, but still much higher than the hadron mass).
Now, the emission of an additional soft gluon
(with one extra power of $\alpha_s$)
gives rise to a large double-logarithmic factor after the
integration over the gluon momentum.
The lowest order amplitude for the
$n$-gluon emission is thus proportional to
\begin{equation}
\frac{1}{n!} \left( \alpha_s \ln ^2 \frac{W^2}{Q_0^2} \right)^n
\end{equation}
where $W$ is the total energy of the hard interaction and $Q_0$ is
the cutoff energy (of the order of the hadron mass),
below which the perturbation theory does not apply.
Although $\alpha_s$ itself becomes smaller at higher energies,
$\alpha_s$ times the double-logarithm becomes larger.
It means that the higher order terms (with larger $n$)
becomes increasingly more important than the lowest order terms.
Therefore, we need to reorganise the perturbation series in order to
establish a well-defined expansion (resummation).

At higher orders, the number of the contributing Feynman diagrams
becomes very large, and each diagram becomes very complicated
because of the presence of the gluon self-coupling.
Consequently, the double-log series takes a very complicated form,
and no obvious pattern (such as the simple exponentiation) seems
likely to be found in it.
One may wonder if there is any asymptotic expression to which
the series approaches at high energies.

There is another question associated with the large logarithms:
their presence would imply
that the cross section would crucially
depend on the infrared cutoff $Q_0$.
Around the cutoff energy $Q_0$,
the effective QCD coupling becomes large,
and the interaction becomes non-perturbative.
At present, we are far from having quantitative understanding of the
non-perturbative hadron formation.

In obtaining the double-logarithmic expression,
we have arbitrarily neglected what is going on
below the energy scale $Q_0$, simply because we do not know about it.
The crucial dependence on $Q_0$, however,
would imply that we could not say anything conclusive
if we did not know the low-energy non-perturbative process.

The key ideas to answer these questions are:

\noindent (1) the angular ordering
in the successive soft-gluon emission;

\noindent (2) the resummation of the large logarithms to all orders:
the modified leading-log approximation; and

\noindent (3) the factorisation of the hadronisation process:
the local parton-hadron duality.

\subsection{the Angular Ordering}

The angular ordering (AO) in the successive soft-gluon emission
was first discovered in the complicated and systematic analysis of
the Feynman diagrams\cite{mue1}.
The essence of the AO, however, can easily be understood
in terms of the coherence
in the emission of a soft gluon at a large angle:
Suppose that an energetic parton decays into a number of nearly
parallel partons, forming a jet,
and a soft gluon is emitted at a large angle from them.
The soft gluon can be emitted from any final line,
and we have to sum all the possible soft-gluon insertions
in order to obtain a gauge-invariant (i.e. physical) result.
After the summation, the soft-gluon emission amplitude factors out
from the jet-production amplitude.
The factorised soft-gluon emission amplitude
depends only on the colour factor of the parent hard parton,
($C_A$ for a gluon, $C_F$ for a quark)
and not on the detailed structure of the jet (Fig.1).
  \begin{figure}
    \vspace{2.0cm}
    \caption{The factorisation of the large-angle soft-gluon emission
    amplitude.
    The solid lines represent partons (quarks of gluons).}
  \end{figure}
Namely, a large-angle soft gluon does not resolve the
detailed transverse structure of the jet,
and can only be emitted coherently from it.
It cannot be emitted from the individual branches
of the parton cascade.
It means that the soft-gluon emission from each individual branch
can take place only at a small angle (AO).

Let us now consider the multiplicity of soft gluons in a gluon-jet
\begin{equation}
<n>_{g-jet}=\int^1_0{\rm d}x\frac{1}{\sigma_T}
\frac{{\rm d}\sigma}{{\rm d}x}\;,
\end{equation}
where d$\sigma$/d$x$ is the the one-gluon inclusive cross section,
and $x$ is the longitudinal momentum fraction of the registered gluon.

Making use of AO, the systematic summation of a large number
of complicated Feynman diagrams can be put in a simple form:
The leading contribution at a given
order in $\alpha_s$ (with the highest power of logarithms) is given by
the configuration

\vspace{0.5cm}

$P \gg k_1\gg k_2\gg \cdots \gg k$ ,

$\theta_1 \gg \theta_2\gg \theta_3\gg \cdots$ .

\vspace{3.0cm}

\noindent The successive factorisation
takes place with this configuration.
The $O(\alpha_s^n)$ cross section
$\sigma_n$ (at tree level) is thus given by
\begin{equation}
\frac{\sigma_n}{\sigma_T}=\prod_i \frac{C_A\alpha_s}{\pi}
\int^{k_{i-1}} \frac{{\rm d}k_i}{k_i}\int^{\theta^2_{i-1}}
\frac{{\rm d}\theta^2_i}{\theta^2_i}
\Theta (k_i^2\theta_i^2-Q_0^2)\;,
\end{equation}
where $\Theta (\xi )$ is the step function:
$Theta (\xi )$=1 for $\xi$>0, and
$Theta (\xi )$=0 for $\xi$<0.
The step function represents the IR cutoff in the transverse momentum.
Because of the presence of the cutoff,
the two logarithmic integrations,
one over the energy, the other over the angle, are linked one another.
This fact makes the explicit form of the double-logarithmic series
to appear very complicated for large $n$.

In terms of $q_i^2$, the virtuality of the $i$-th gluon, and $x_i$, its
longitudinal momentum fraction to the $(i-1)$-th gluon,
(4) can be rewritten as
\begin{equation}
(4)=\prod_i \frac{C_A\alpha_s}{\pi}
\int^{x_{i-1}q^2_{i-1}} \frac{{\rm d}q^2_i}{q^2_i}
\int^1_{\frac{q_i^2}{x_{i-1}q^2_{i-1}}}
\frac{{\rm d}x_i}{x_i}\Theta (x_iq_i^2-Q_0^2)\;.
\end{equation}
$x_iq_i^2$ in the argument of the step function
is the transverse momentum
square of the $i$-th soft gluon.
Note that the kinematical upper bound of $q_i^2$
is given by the transverse momentum square of the $(i-1)$-th gluon.
This fact is important because it allows us
to write the sum of the perturbative series
in the form of a compact integral equation in the next subsection.

\subsection{the Modified Leading-Log Approximation}

In the case of the $e^+e^-$-annihilation, an energetic $q \bar q$
is first produced.
The multiplicity (of gluons) is given by
\begin{equation}
<n>_{e^+e^-}=\frac{2C_F \alpha_s}{\pi} \int_0^{W^2} \frac{\d q^2}{q^2}
\int^1_{q^2/W^2}\frac{\d x}{x}M_g(xq^2)\;,
\end{equation}
where $W$ is the total centre-of-mass energy, and
\begin{eqnarray}
M_g(q^2)&=&\sum_{n=0}^{\infty}M_g^{(n)}(q^2)\;,\\
M_g^{(0)}&=&1\;,\nonumber\\
M_g^{(n)}(q^2) &=& \frac{C_A \alpha_s}{\pi} \int_0^{q^2}
\frac{\d q_1^2}{q_1^2}
\int_{q_1^2/q^2}^1 \frac{\d x_1}{x_1} \, \frac{C_A \alpha_s}{\pi}
\int_0^{x_1 q_1^2} \frac{\d q_2^2}{q_2^2} \int_{q_2^2/x_1q_1^2}^1
\frac{\d x_2}{x_2} \times \cdots
\nonumber\\
\nonumber\\
& & \times \frac{C_A \alpha_s}{\pi} \int_0^{x_{n-1}q^2_{n-1}} \frac
{\d q_{n}^2}{q_{n}^2} \int_{q_{n}^2/(x_{n-1}q_{n-1}^2)}^1
\frac{\d x_n}{x_n}  \quad .\nonumber
\end{eqnarray}
The IR cutoff (in the transverse momentum square)
is made implicit in (7),
in order to avoid the notational complexity.

Now the series (7) can be represented formally by the integral equation
\begin{equation}
M_g(q^2)=\frac{C_A\alpha_s}{\pi} \int_0^{q^2} \frac{\d q_1^2}{q_1^2}
\int_{q_1^2/q^2}^1 \frac{\d x}{x} M_g(xq_1^2)+1\;.
\end{equation}
It can easily be checked that the iteration of the equation (8) gives
the series (7) explicitly.

Now, because $M_g(q^2)$ increases rapidly at high energies (in fact,
it diverges as the IR cutoff vanishes),
the inhomogeneous term on the rhs of (8) becomes negligibly small
compared with the homogeneous term.
The equation can be solved easily if we neglect the inhomogeneous term:
Because the integration kernel does not include any dimensional
parameter, the solution is given by a power of $q^2$
\begin{equation}
M_g(q^2) \propto (q^2)^{\gamma}
\end{equation}
and the integration simply counts the power $\gamma$:
\begin{eqnarray}
\int_0^{q^2} \frac{\d q_1^2}{q_1^2} \int_{q_1^2/q^2}^1 \frac{\d x}{x}
(q_1^2)^{\gamma}
&=& \int_0^{q^2} \frac{\d q_1^2}{q_1^2} \left\{ \frac{1}{\gamma}
(q_1^2)^{\gamma}-\frac{1}{\gamma} \left( \frac{(q_1^2)^2}{q^2}
\right) ^{\gamma} \right\}
\nonumber\\
&=&\frac{1}{2\gamma^2} (q^2)^{\gamma}
\end{eqnarray}
We therefore obtain
\begin{equation}
(q^2)^{\gamma} = \frac{C_A\alpha_s}{\pi}\frac{1}{2\gamma^2}
(q^2)^{\gamma} \quad \rm{or} \quad
\gamma^2= \frac{C_A \alpha_s}{2\pi}
\end{equation}

The anomalous dimension $\gamma$ is proportional to $\sqrt{\alpha_s}$.
Namely, the solution (9) cannot be expanded in powers of $\alpha_s$.
It implies that the solution is an asymptotic form in the large
multiplicity limit.
In fact, when $W^2/Q_0^2$ is of the order one, the solution of the
inhomogeneous equation depends on the IR cutoff $Q_0$.
As the multiplicity increases at higher energies, the inhomogeneous
term becomes negligible,
and the solution approaches the asymptotic form.

The asymptotic from (9) does not include the IR cutoff explicitly.
It implies that the $Q_0$-dependence factors out
in the normalisation constant in the solution.
The normalisation constant is not determined by a homogeneous
linear integral equation.

This factorisation of the IR singularity is somewhat analogous to the
well-known factorisation of the ultraviolet (UV) divergence.
In a renormalisable field theory, UV-divergences are factored out into
the normalisation of a limited number of physical parameters,
which are fixed by the observed quantities.
Namely, in such a theory, the observed phenomena are described without
referring to what is going on at short distance (near the UV cutoff).
In the case of the IR renormalisation discussed here, the multiplicity
is described without referring to what is going on at long distances
(near the IR cutoff).

The integral equation (8) can be combined with the more conventional
leading-log approximation\cite{gri},
in which the collinear signle-logarithms
are summed to all orders.
It is done by adding in the rhs of (8) the non-soft integration kernel
evaluated in the light-cone gauge.
The kernel is then essentially identical
with the familiar Altareli-Parisi
splitting function.
This formalism includes both soft and collinear singularity correctly
at the next-to-leading order (i.e. $O(\gamma^2)=O(\alpha_s)$),
and is called the modified leading-log approximation\cite{mue3}.

\subsection{Local Parton-Hadron Duality}

So far, we have discussed only the multiplicity of gluons.
What is the relation between the gluon multiplicity,
calculated in MLLA, and
the hadron multiplicity observed in the experiments?

The hypothesis of the local parton-hadron duality
(LPHD)\cite{dok2} states
that the hadron formation takes place locally
so that only a limited amount of momentum
(i.e. of the order of $Q_0$, and independent of the total energy $W$)
is exchanged during the hadronisation.
Then the hadron multiplicity would be
proportional to the gluon multiplicity,
the proportionality constant being independent of the jet energy.
The particle flow pattern at the parton level would be the same as the
particle flow at the hadron level except for the normalisation.

The hypothesis is based on the observation in the perturbation theory,
in which the soft-gluon emission causes the screening of the colour
charge so that colour-singlet clusters of a limited invariant mass are
formed after the partonic branching\cite{ama}.
It seems natural to assume that the hadronisation takes place mostly
within each colour-singlet cluster (\lq\lq preconfinement").

Because of the lack of our theoretical knowledge of the hadronisation,
there is no rigorous proof of this hypothesis.
We have to examine it by extensively comparing the calculation at the
parton level (MLLA) with the experimental hadronic data.

Under LPHD, we can reinterpret the expression (6) with (9) as that for
the hadron multiplicity itself.
Comparing the prediction including next-to-leading order
correction\cite{mue3} with the experimental data, we find no
significant deviation in a wide range of energy (10-91GeV).
Other phenomenological model
for the multiple hadronisation, such as LUND
string fragmentation model combined with $O(\alpha_s)$ matrix element,
failed to reproduce the observed rise of the multiplicity as $\ln W^2$
increases.
The sharp increase of the multiplicity is characteristic to MLLA,
in which the multiplicity of soft gluons increases
through the branching process with the gluon self-coupling.
The same increase observed
in the experiments suggests that the multiplicity
increase of the gluons is directly reflected
in the hadron multiplicity.

Another remarkable success is found
in the hadron energy spectrum\cite{opa2}.
Because of the successively softer gluon emission,
there are more soft gluons than harder ones.
The spectrum thus increases as $\ln (1/x)$ increases
(i.e. the energy of a gluon decreases).
There is, however, a lower bound $Q_0$
for the transverse momentum of the gluon emission.
A very soft gluon, therefore, can be emitted only at a large angle.
Owing to the coherence discussed above,
the large angle soft gluon does not see the parton proliferation
which takes place mostly at small angles.
Therefore, the multiplicity of the softest gluons remains unchanged
even when the total energy $W$ is increased.
The resulting spectrum takes a
nearly symmetric bell-shape in $\ln (1/x)$.
These features of the gluon spectrum are found in the hadron spectrum.
The agreement of the theoretical gluon spectrum and the
experimental hadron spectrum is strikingly good.

\section{Tests of the Theory}

Although the MLLA with the hypothesis LPHD is
successfully applied to the multiplicity in the e$^+$e$^-$-annihilation
and to the hadron spectrum,
it is at present not necessarily as successful for other quantities.
Particular difficulties are felt in the shape
of the multiplicity distribution and in the multiplicity ratio.

In the following, I'll discuss these and other related quantities,
clarifying what kind of problems are in it.

The quantities we shall discuss below are:

\noindent (1) the multiplicity distribution;

\noindent (2) the multiplicity ratio
between a gluon-jet and a quark-jet;

\noindent (3) the multiplicity correlation; and

\noindent (4) the angular distribution of hadrons in a jet.

\subsection{The Multiplicity Distribution}

When we draw the multiplicity distribution in a scaled variable
$\zeta =n/\langle n\rangle$,
where $n$ is the number of hadrons in an event,
and $\langle n\rangle$ is its average,
the distribution observed at high energies appears
to be independent of the total energy $W$\cite{ale1}.
The energy-independence of the scaled distribution
is called KNO scaling\cite{kob} (Fig.2a).
It was shown\cite{tes1} that the KNO scaling
in the $e^+e^-$ annihilation is the consequence of the scale invariance
(asymptotic freedom) in the presence of the gluon self-coupling.
It is thus the manifestation
of the essence of the non-abelian gauge theory.
With MLLA, we can evaluate the asymptotic shape of the distribution
and its higher order corrections systematically.
The problem, however, is that the asymptotic shape of the distribution
evaluated at the leading order is too far from the observed one.
  \begin{figure}
    \vspace{2.0cm}
    \caption{(a) The KNO scaling of the multiplicity distribution [8].
    The solid curve is the result of the LUND 7.2 parton shower model
    with parameters tuned at $W$=91.25GeV.
    (b) The squared dispersion of the scaled multiplicity distribution.
    The asymptotic value $r=3/8$ is compared with the evaluation by the
    improved MLLA [12].
    The data of the multiplicity correlation are also shown by crossed
    points.}
  \end{figure}

The squared dispersion of the scaled distribution, for example, is
\begin{eqnarray}
D^2_{\rm theory}&=&\frac{<(n-<n>)^2>}{<n>^2}=<\zeta^2>-1
=\frac{C_A}{6C_F}\nonumber\\
&=&\frac{3}{8}=0.375\; ,
\end{eqnarray}
Its observed value, on the other hand, is
\begin{equation}
D^2_{\rm observed}\approx 0.08{\rm -}0.09 \;\;
({\rm at}\; W=28{\rm -}91{\rm GeV})\; ,
\end{equation}

The next-to-leading order correction ($O(\sqrt{\alpha_s})$)
to the squared dispersion was evaluated and was found to be so
large that it almost cancels the leading order value\cite{mal}.
It implies that at the current energies
the higher order corrections are systematically large.
The perturbative series in powers of $\sqrt{\alpha_s}$
violently oscillates at higher orders.
We need to sum up the large contributions to all orders
to reorganise the perturbation series (re-resummation).

Such redefinition of the perturbation series was done in \cite{cuy}.
The result, as is shown in Fig.2b,
is much closer to the observed values,
and we do not have the problem
of the divergently oscillating higher orders.

To be exact, the quantity which is calculated in MLLA
is the multiplicity correlation,
$\langle n(n-1)\rangle /\langle n\rangle^2-1
=\langle\zeta^2\rangle-1-1/\langle n\rangle$,
in stead of the squared dispersion of the scaled distribution,
$\langle\zeta^2\rangle-1$.
(The former is sometimes called the second \lq\lq factorial moment".)
The difference between them is $1/\langle n\rangle$.
It does not appear at any finite order in the power series
in $\sqrt{\alpha_s}$ (the non-perturbative correction),
and we have not distinguished the two quantities so far.
The magnitude of the difference, however,
is not necessarily negligible at our current energies,
because the quantity in question happens to be small.

The difference between the experimental data
of the multiplicity correlation
and its theoretical evaluation is still large even after
the improvement of the approximation.
There are three possibilities for the cause of the discrepancy:

\noindent (i) yet uncalculated higher-order corrections;

\noindent (ii) non-perturbative effects; or

\noindent (iii) the LPHD might be incorrect or have to be modified.

If the third possibility should be the case,
it would force us to reexamine the theoretical framework
as a whole to analyse the multiple hadroproduction in terms of
the perturbative QCD.
It is therefore significant to evaluate the first two within the
framework of MLLA+LPHD.

The first one, however, is technically hard to
evaluate at this moment.
The second one is present because
$Q_0/W$ is not negligibly small at our current energies
(we call such a correction the finite-$Q_0$ effect).
In this case, it is of the order of magnitude of $1/\langle n\rangle$.

The possible significance of this effect may be understood as follows:
The positive contribution to the multiplicity correlation comes from
the gluon self-coupling, as is mentioned above.
If gluons were emitted from the initial $q\bar q$-pair independently,
as is the case of the abelian gauge theory,
the positive contribution would vanish.
(In fact, the recoil effect
due to the energy-momentum conservation gives
negative contribution to the correlation\cite{cuy}.)
Now suppose that the transverse momentum cutoff $Q_0$ is around 1GeV.
Then only several gluons would be emitted from the initial quark or
antiquark at TRISTAN or LEP energies.
The emission is mostly at a small transverse momentum.
There is not much chance
for the gluon emitted at a small transverse momentum
to undergo further branching {\em because of the angular ordering}.
If this be the case, the correlation can be very small
(or can even be negative because of the recoil effect).

One of the ways to examine the presence of the finite-$Q_0$ effect
is to go to much higher energies
so that $1/\langle n\rangle$ may be negligibly small.
An alternative way, which may be tried at the TRISTAN energies,
is to change $Q_0$ artificially:
It can be done by measuring the distribution of the
{\em multiplicity of jets},
defined by $k_T$ (so-called Durham) algorithm\cite{dur}
(with varying $k_T$-cut).
We then compare the results with the theoretical calculation,
which can be done by a numerical integration of the integral
equation of the multiplicity moment\cite{cuy}.
An alternative way for the numerical integration is to make use of
the shower MC simulation program {\em at the parton level} with varying
cutoff size.
The theoretical study is now underway, which will give more
insight on this problem.

\subsection{The Multiplicity Ratio}

Another serious discrepancy between the theory and the experiments
was found in the ratio between the multiplicity of hadrons
in a gluon jet and that in a quark jet.
At the leading order of MLLA, the multiplicity ratio is
identical to the ratio of the colour charges of the
corresponding hard partons ($C_A=3$ for a gluon, $C_F=4/3$ for a quark).
\begin{equation}
r=\frac{<n>_{\rm gluon-jet}}{\rm <n>_{quark-jet}}=
\frac{C_A}{C_F}=\frac{9}{4}=2.25\; .
\end{equation}

The ratio has been measured using three-jet events in the
$e^+e^-$-annihilation,
identifying one of the three jets as the gluon-jet.
It was found much lower than the above leading-order value.
For example, OPAL collaboration gave \cite{opa1}
\begin{equation}
r={\langle n\rangle_{\mbox{\tiny g-jet}}\over \langle
n\rangle_{\mbox{\tiny q-jet}}}
=1.267\pm 0.043\pm 0.055({\rm syst.})
\end{equation}
(including neutral particles) and
\begin{equation}
r_{\mbox{\tiny CH}}={\langle n_{\mbox{\tiny CH}}
\rangle_{\mbox{\tiny g-jet}}
\over \langle n_{\mbox{\tiny CH}}\rangle_{\mbox{\tiny q-jet}}}=
1.326\pm 0.054\pm 0.073({\rm syst.})
\end{equation}
(charged particles only).
(For earlier experiments, see \cite{rat}.)

The $O(\sqrt{\alpha_s})$ correction in MLLA reduces the leading order
value (14) by about 10 percent \cite{mue2}.
The correction is far too small to explain the discrepancy from the
experimentally observed value.
(Next-to-next-to-leading order does not change the situation
\cite{guf}.)

It should be noted here that the theoretically calculated quantity
is not necessarily identical to the experimentally measured one.
In the theoretical calculation,
$\langle n \rangle_{\mbox{\tiny g-jet}}$ is defined as a half of
the total multiplicity from the two hard gluons created by a
gauge invariant gluon source (for example, $F_{\mu \nu}^2$,
where $F_{\mu \nu}$ is the field strength of gluon).
Experimentally, on the other hand, {\em well-separated}
three-jet events are selected, and the number of hadrons
in each isolated angular region is compared one another.
The two quantities are not identical in their definition.
Particularly, the value of the latter
(experimental one) depends, in general,
on the event selection criteria.

It is, therefore, desirable to make the measurement of the ratio
which does not make use of the specific event selection.
The bias-free measurement, however,
is not easy because a gauge-invariant
two-gluon source is hard to prepare.

The use of the value of thrust is proposed in \cite{kim}
to detect the hard-gluon emission (in place of identifying
the presence of well-separated three jets).
Thrust $T$ is defined, in the c.m. $\!$frame, by
\begin{equation}
T = {\rm max}\left\{\frac{\sum_{i} \mid \vec{p}_i\cdot \vec{n}
\mid}{\sum_{i} \mid \vec{p}_i \mid }\right\}~~
(\vec{n}^2 = 1)
\end{equation}
(the direction of the three-vector $\vec{n}$ is chosen
to maximise the rhs).
$T=1$ would imply that all the particles are either parallel
or antiparallel to $\vec{n}$.

If $T$ is far from one, it implies that a hard gluon is emitted.
At $T=2/3$, the only configuration at the lowest order ($O(\alpha_s)$)
is 3-fold symmetric (a symmetric 3-jet).
In fact, the $O(\alpha_s)$ contribution has a vanishing phase space
at $T$=2/3, and $O(\alpha_s^2)$ matrix element gives the
dominant contributions around $T$=2/3.
Because of the presence of the soft and collinear singularities,
one of the two gluons is likely to be much softer than the other.
It is also likely to be collinear to one of the three hard partons.
Consequently, the dominant configuration
at $O(\alpha_s^2)$ (and higher)
is also more or less 3-jet-like.

The emitted hard gluon causes an increase of the multiplicity
from the case of the hard $q\bar{q}$-pair only.
It is therefore natural to associate the multiplicity increase
near $T$=2/3 to the additional third jet (the gluon-jet).
In fact, we can define
$\langle n\rangle_{gluon}$ by
\begin{eqnarray}
\langle n\rangle_{gluon}&=&\langle n\rangle_{T=2/3}
-\langle n\rangle_{W^2/3}\;,\\
\langle n\rangle_{quark}&=&\frac{\langle n\rangle_{W^2/3}}{2} \; ,
\end{eqnarray}
where $\langle n\rangle_{W^2/3}$ is the average multiplicity of the
e$^+$e$^-$-annihilation at the centre-of-mass energy $W/\sqrt{3}$.
($W/\sqrt{3}$ is equal to the centre-of-mass energy of the $q\bar q$
in the symmetric 3-jet configuration.)
It was shown in \cite{kim}, by an explicit calculation in
MLLA at the next-to-leading order, that the ratio
$R(1-T)=\langle n\rangle_{gluon}/
\langle n\rangle_{quark}$ at $1-T$=1/3 is
equal to the multiplicity ratio $r$ (defined with use
of a gauge-invariant two-gluon source \cite{mue2}).
\begin{eqnarray}
R(1/3)&=&r+O(\alpha_s)\;, \\
R(1-T)&\equiv &\frac{\langle n\rangle_{gluon}}{\langle n\rangle_{quark}}
\; ,\nonumber\\
r&=&\frac{C_A}{C_F}\left\{1-\frac{\gamma_0}{3\sqrt
{\ln W^2/\Lambda_{QCD}^2}} \left( 1-\frac{N_f}{C_A}
(2\frac{C_F}{C_A}-1)\right) \right\}\; .\nonumber
\end{eqnarray}

In this method, we do not identify and separate the gluon jet.
The quantity $\langle n\rangle_{gluon}$
defined by (18), therefore, may not
necessarily be identifiable as the multiplicity of
hadrons {\it in a gluon jet}
(if the energy is not asymptotically high).
Nevertheless, this method has an obvious advantage:
What is experimentally
measured is the same quantity defined and evaluated in the theory.
The comparison between the experimental data and the theoretical
prediction is thus direct and unambiguous.
If we obtained reasonable agreement between them,
we would be able to claim
that the multiplicity from the hard-gluon emission
seem to be consistent with the theory,
and that the discrepancy between the theoretical value of $r$
and its experimental value obtained by selecting well-separated
3-jet events be likely because of the bias introduced by the event
selection criteria\footnote{
There may may be other reasons, such as the finite-$Q_0$ effect,
which I shall mention shortly
in connection with the sub-jet multiplicity.}.

$R(1/3)$ was measured by TOPAZ and AMY.
The results will be reported later in this session\cite{liu}\cite{nak}.
The values they obtained are somewhat lower than
(but not very far from) the theoretical value of $r$.
It should be noted that the two quantities, $R(1/3)$ and $r$,
are not identical in their definitions
and the uncalculated difference is
of the order of magnitude of $\alpha_s$.
It should be also noted that the experimental value
of $\langle n\rangle_T$ at $1-T=1/3$ is
obtained by extrapolating the data at lower values of $1-T$,
assuming its linear dependence on ln($1-T$).
Taking these uncertainties into account,
I'd say that the observed values
of $R(1/3)$ seem consistent with the theoretical value.
Further improvement in the theoretical evaluation is desirable
if we want a more definite conclusion.

\vspace{0.5cm}

\noindent {\bf Related Quantities}

\vspace{0.2cm}

In order to avoid the unevaluated influence of the event selection
criteria in the measurement of the multiplicity in a gluon-jet,
Catani et al.\cite{cat} proposed to measure the ratio
of sub-jet multiplicity in the 3-jet events
and in the 2-jet events defined by the Durham algorithm\cite{dur}.
The measurement of this quantity
at TRISTAN and comparison with the theory
will be presented later in this
session\cite{beh}(see also \cite{ale2}).
They found discrepancies from the theory for small values of
the cutoff $k_T$ which defines the separation of the subjets
(below $k_T\approx$2GeV).

We may interpret the result in the following way:
For the vanishing cutoff $k_T$ value, the ratio between
the sub-jet multiplicities becomes identical to the ratio between
the hadron multiplicities.
In this limit, the sub-jet multiplicity for 3-jet events
is an analogous (though not necessarily identical)
quantity to $\langle n\rangle_{T=2/3}$.
We expect the latter, at the next-to-leading order of MLLA,
to be around 1.6 times $\langle n\rangle_{e^+e^-}$
at the current energies \cite{kim}.
The observed value of the sub-jet multiplicity ratio
in the vanishing cutoff $k_T$ limit is considerably lower than 1.6.
On the other hand, according to a shower Monte Carlo simulation,
it seems to approach the value somewhere around 1.6 at higher energies.
This fact suggests that the observed discrepancy
between the theoretical calculation and the experimental data
of the sub-jet multiplicity ratio for small cutoff $k_T$ values
may well be a finite-$Q_0$ effect.
If this is the case, the result is very interesting
because we have so far very limited information on the size of $Q_0$.

The quantities calculable in MLLA are normally insensitive to $Q_0$
at high energies, and we have to go to the soft limit at which
MLLA can no more be applied.
The analysis of the hadron spectrum suggests that the cutoff scale
may be far below 1GeV, but the evidence was rather indirect.
The data of the sub-jet multiplicity may help us to determine
the size of $Q_0$.
Careful analysis, however, is required in such determination because
there is the possibility that the uncalculated higher order corrections
in MLLA, rather than the non-perturbative process,
may be responsible to the discrepancy.

\vspace{0.3cm}

We may also try to estimate the hadron multiplicity in a gluon jet
in a more direct way, but without
requiring that the event consists in {\em well-separated} three jets.
Let us use the value of thrust again.

\noindent (i) Take events with the value of $1-T$ not small,
but not as large as 1/3, say 0.10 to 0.15.
Divide hadrons into the two hemispheres divided by the plane
perpendicular to the thrust axis.
In most cases, the invariant mass of one hemisphere is much larger
than the other.
A hard gluon is most likely to be emitted in the hemisphere
with larger invariant mass.

\noindent (ii) Take all the hadrons in this hemisphere,
and go to the centre-of-mass frame of them.
Evaluate the thrust axis of these hadrons.
Select the events with the new thrust axis at large angles from the
boost direction (identical to the direction of the old thrust axis).
The system has the centre-of-mass energy about $\sqrt{1-T}W$
($\approx$20GeV for TRISTAN experiments),
and is likely to be more or less two-jet like,
one of the jet being from
a hard gluon.

\noindent (iii) Measure the asymmetry in the hadron multiplicity
along the new thrust axis.
Most naively, the ratio of the multiplicity between two sides would be
equal to the multiplicity ratio $r$.
There are corrections to it,
which are calculable in the perturbative QCD.
Measuring this quantity, and comparing the results with the
theoretical evaluation would give further information on the
multiplicity ratio.

\subsection{Multiplicity Correlation}

One of the most remarkable predictions of MLLA
is the \lq\lq drag effect" \cite{azi}
(or \lq\lq the coherence of the second kind"):
The emission of a hard gluon from the $q\bar q$-pair
causes the change in the colour flow,
and this change is reflected in the modification
of the global particle flow pattern.
As a consequence, the particle density at the direction opposite
to the hard gluon is depleted,
compared with the case of the absence of the gluon.

This phenomenon was first predicted in the context of the
old string model of the multiple hadroproduction combined with the
$O(\alpha_s)$ cross section in QCD.
In this string model, the hadronisation process
(and not the partonic process) is assumed to be responsible
for the multiplicity increase at higher energies.
Although this picture itself is incorrect as the model of the
multiple hadroproduction,
MLLA gives in effect the same phenomenon.
The observation of the drag effect was reported
in a number of experiments\cite{jad} (Fig.3a).
  \begin{figure}
    \vspace{2.0cm}
    \caption{(a) The interjet asymmetry in the three-jet events[25].
    The particle density in the direction opposite to the gluon
    jet is smaller.
    (b) The azimuthal correlation:
    The AMY data are compared with the Mote Carlo simulations[28].}
  \end{figure}

Some argue, however, that other kinematical and stochastic
effects may cause similar effects.
Only the quantitatively precise agreement between the theory (MLLA)
and experiments would provide
the conclusive evidence of this phenomenon.
Unfortunately, the results of the measurements
with the use of the well-separated three-jet events
may depend on the jet-selection criteria,
and the direct quantitative comparison with the theoretical
calculation is not easy.

In order to test the drag effect
without being influenced by the jet-selection procedures,
Dokshitzer et al.\cite{dok3} proposed to measure the multiplicity
correlation (the energy-multiplicity-multiplicity correlation).
They defined the multiplicity correlation $C(\phi)$ by
\begin{equation}
C(\phi)=\frac{C_{EMM}(\eta_{min},\eta_{max},\phi)C_E}
{\left[C_{EM}(\eta_{min},\eta{max})\right]^2}\;,
\end{equation}
where $C_{EMM}$ is the energy-multiplicity-multiplicity correlation
\begin{eqnarray}
C_{EMM}(\eta_{min},\eta_{max})&=&\frac{1}{\sigma}
\int E_i{\rm d}E_i{\rm d}E_j{\rm d}E_k
\int^{\eta_{max}}_{\eta_{min}}{\rm d}\eta_j{\rm d}\eta_k \nonumber\\
&&\times\int^{2\pi}_0 {\rm d}\phi_j{\rm d}\phi_k
\delta(\phi -\phi_j +\phi_k)\frac{{\rm d}\sigma}
{{\rm d}E_i{\rm d}E_j{\rm d}E_k{\rm d}\eta_j{\rm d}\eta_k
{\rm d}\phi_j{\rm d}\phi_k} \;,
\end{eqnarray}
$C_{EM}$ is the energy-multiplicity correlation:
\begin{equation}
C_{EM}(\eta_{min},\eta_{max})=
\frac{1}{\sigma}\int E_i{\rm d}E_i{\rm d}E_j
\int^{\eta_{max}}_{\eta_{min}}{\rm d}\eta_j
\int^{2\pi}_0 {\rm d}\phi_j
\frac{{\rm d}\sigma}{{\rm d}E_i{\rm d}E_j{\rm d}\eta_j\phi_j}\;,
\end{equation}
and
\begin{equation}
C_E=\frac{1}{\sigma}\int
E_i{\rm d}E_i\frac{{\rm d}\sigma}{{\rm d}E_i}\;.
\end{equation}
In taking the ratio to obtain $C(\phi)$, the normalisation constant in
the multiplicity cancels out between the numerator and the denominator.
The result is thus free
from the ambiguity in the normalisation constant,
which cannot be determined in the perturbation theory.
The most heavily weighted directions of the energy-weighted particle
(the directions of the energy flow),
labeled by the subscript $i$
in the integration on the rhs of (22) and (23),
can be identified as the jet directions.
$C_{EM}(\eta_{min},\eta_{max})$ is the multiplicity in the
longitudinal rapidity interval (defined relative to the jet direction),
while $C_{EMM}(\eta_{min},\eta_{max},\phi)$ is the multiplicity
correlation in the same interval.
$C(\phi)=1$ would imply the absence
of the correlation between two hadrons.

At $O(\alpha_s^2)$ (the lowest non-trivial order),
we obtain the back-to-back correlation as
\begin{equation}
C(\pi)=\frac{7}{16}=0.4375\;.
\end{equation}
This result holds even
after the resummation (at the leading order in MLLA).
It is owing to the angular ordering:
Because the emission of a soft gluon from the primary gluon
(emitted directly from the initial $q\bar q$-pair) is at a small angle,
it does not modify the back-to-back correlation.

Unfortunately, if we add the next-to-leading order
correction \cite{dok4},
the correlation becomes close to one (no correlation) at our energy:
\begin{equation}
C(\pi)\approx 0.93\;({\rm at\; the\; next\; to\; leading\; order}).
\end{equation}
The situation is somewhat similar
to what we have encountered in the case of
the squared dispersion of the multiplicity distribution\footnote{
Note also the difference between the two cases.
In the squared dispersion,
the gluon self-coupling is responsible for the positive contribution,
and thus the small angle correlation is important,
while the drag effect concerns itself with the large angle correlation,
and the gluon self-coupling does not contribute to it at the
leading order.}.
Because the next-to-leading order correction almost cancels the
negative correlation at the leading order,
we cannot predict anything quantitatively
in the power series in $\sqrt{\alpha_s}$ at the current energies.
We need some improvement of the perturbative expansion,
as was done in the case of the multiplicity distribution.

The correlation was measured at LEP and TRISTAN \cite{opa3}.
They obtained the values which lie between the two values (25) and (26),
as one might have expected:
\begin{equation}
C(\pi)_{observed}\approx 0.78\;.
\end{equation}

Interestingly enough, they also found that they could
reproduce the observed correlation by the shower Monte Carlo
simulation in which the above drag effect (coherence of the
second kind) was not implemented.
It implies that what was observed in the back-to-back correlation
may not necessarily have been the drag effect itself.
What then could cause the negative correlation without the drag effect?

In the case of the squared dispersion of the multiplicity distribution,
it was found that the recoil effect gives rise to a large negative
correlation at the current energies\cite{cuy}:
The emission of a gluon,
when it gives a significant contribution to the multiplicity,
carries out a non-negligible fraction
of the energy of the parent parton.
It would reduce the multiplicity
from a gluon emitted later from the same
parent parton.

In addition to the recoil effect,
the finite-$Q_0$ effect might as well be significant,
and should also be analysed carefully.

Without the evaluation of the recoil effect,
as well as other possibilities
which might affect the back-to-back correlation,
we cannot tell how much of the observed negative correlation is
due to the drag effect.

Another interesting point was found when
they ran a shower Monte Carlo program, in which the angular ordering
(the coherence of the first kind) was not implemented,
but otherwise identical to the LUND shower MC.
At the parton level, they found an almost
vanishing back-to-back correlation.
The negative correlation, however, reappeared among the hadrons
which are generated from the partons produced
without angular ordering (Fig.3b).

The reason for the reduction
of the negative correlation at the parton level
is because the back-to-back emission
of a soft gluon from the primary gluon
is allowed in the absence of the angular ordering.
The possibility of back-to-back emission
is suppressed again {\em at the hadron level} in the following way:
In order to produce hadrons in the LUND shower MC model,
strings are attached to the pairs of partons.
When a string is attached between the two back-to-back gluons,
one of which is much softer than the other,
its centre-of-mass momentum is shifted toward the direction of the
first (less soft) gluon.
Therefore, the hadrons are mostly generated on one side.
This suppression of the back-to-back emission causes
the negative correlation at the hadron level.
The effect of the string hadronisation thus mimics
the role of the angular ordering at the parton level
in the back-to-back correlation.

This is an example that the results of a MC model,
whose partonic process is not justified theoretically,
could be modified by the model-dependent hadronisation process
to give approximately correct simulation results for some quantities.
Such a simulation, however, cannot be \lq\lq universal".
In fact, they found that the MC simulation without the angular ordering
(but with the string hadronisation) fails to reproduce the
correlation for the angles other than in the back-to-back region.

\subsection{Angular Distribution of hadrons in a Jet}

In the previous section, I have discussed the particle flow
(the angular density of the particles) in the inter-jet region.
Let us now turn to the particle flow in the jet direction.

The angular distribution of hadrons
in a jet was evaluated in \cite{smi}.
The calculation was based on the fact that at the leading order of MLLA
the emission angles in the successive soft-gluon emission
are strongly ordered so that the direction of the registered hadron
can be identified with the direction of the primary gluon,
which is directly emitted from the initial $q\bar q$-pair.
It was found that in the small angle region
the recoil from the multiple soft-gluon emission
from the initial $q\bar q$-pair gives a significant modification
of the shape of the distribution evaluated in MLLA.
The results agree reasonably well with the TASSO data from 14-34GeV.

It is desirable, however, to improve the approximation
in order to examine the theory more precisely.
Particularly, the assumption of the strong angular ordering
may seem rather a crude approximation.
The direction of the primary gluon is in fact not identical to
the angle of the registered particle at higher orders,
and we have corrections due to the difference.

In order to evaluate the correction, we have to analyse the emission
of two gluons at comparable angles.
When both angles are large, the approximation of the angular ordering
needs modification.
Note that such a modification should be necessary by any means
because the order among angles is not a Lorentz invariant quantity.
(The correction is called the \lq\lq dipole correction".)

The correction was analysed in detail in \cite{kim2}.
It was found that the boost invariance is restored when one
adds the dipole corrections to the results
obtained under the angular ordering.

An improved evaluation
of the angular distribution was done in \cite{kim3}.
The precise comparison of the theoretical calculation
with the experimental data at TRISTAN will provide us further
information on the jet physics.
Of particular interest would be the size of the cutoff $Q_0$,
on which the particle density in the jet direction weakly depends.

Another interesting aspect analysed in \cite{kim3} is the mass effect.
When the energetic parton is a $b$-quark,
the mass-singularity in the direction
of the jet is considerably reduced,
because the $b$-quark mass is not negligibly small compared with the
jet energy in the TRISTAN experiments.
(Furthermore, the soft-gluon emission amplitude at the lowest order
vanishes exactly at the direction of the $b$-quark\cite{mar}.)
The reduction of the multiplicity at small angles (Fig.4a),
however, is compensated by the hadrons from the $B$-meson decay.
Consequently, the difference
in the angular distribution of hadrons between
a $b$-quark jet and a light-quark jet is not so prominent (Fig.4b).
  \begin{figure}
    \vspace{2.0cm}
    \caption{The prediction of the distribution of hadrons in a jet
    at $W$=58GeV [31].
    A light-quark jet (solid curve) is compared
    with a $b$-quark jet.
    (a) The dashed curve shows the distribution of the direct hadrons
    in a $b$-quark.
    The recoil from the multiple soft-gluon emission is not included.
    (b) Hadrons from the $B$-meson decay
    are included for the $b$-quark jet.
    The recoil is taken into account so that the sharp peak in (a)
    in the distribution for a light-quark jet is smeared.}
  \end{figure}

In fact, it is not impossible to distinguish the direct hadrons
(i.e. formed before the $B$-meson decay) from those after the decay,
by precisely identifying the decay vertex.
The measurement of the angular distribution of the former will give
very distinctive shape from the light-quark jet.

\vspace{0.5cm}

\noindent {\bf Jet width}

\vspace{0.2cm}

A particularly interesting quantity
associated with the angular distribution
is the \lq\lq angular width" of a jet $W_A$ defined by
\begin{equation}
W_A\equiv <(1-\cos\theta)>\;,
\end{equation}
where the angle $\theta$ is the angle of each registered hadron
relative to the thrust axis.
The average is over one of the hemispheres
divided by the plane perpendicular to the thrust axis.
One may evaluate this quantity event by event, as a characteristic of
the event shape.
Or one may average it over all the hadronic events in the experiment.
In the following, we discuss the averaged one.

It is the characteristic of the average event shape which is determined
by the soft-gluon radiation, and thus is governed by MLLA.
In fact, at the leading order, we find
\begin{equation}
<W_A>=\gamma +O(\gamma^2)\;,
\end{equation}
where $\gamma$ is the anomalous dimension of the multiplicity
(see eqs(9)(11)).

The predictions in MLLA are, in general, unique except for the
arbitrariness of the energy-independent
overall normalisation constant and
the dependence on the cutoff $Q_0$ (the latter vanishes asymptotically).
In the jet width $W_A$, the normalisation constant does not appear.
It is also insensitive to the cutoff $Q_0$.
It is therefore particularly suited to the precise quantitative test
of the theory, MLLA+LPHD.

The next-to-leading order correction ($O(\gamma^2)$) was calculated
in \cite{kit}.
When the next-to-next order correction is calculated,
this quantity can be used for the precision measurement
of the QCD mass scale $\Lambda_{\overline{MS}}$.

\section{Concluding remarks}

The theoretical framework to analyse the multiple hadroproduction,
MLLA with LPHD, has to be extensively tested against experimental data.
In order to be able to apply it to the reaction at higher energies,
it is crucial to improve the approximation.
The refinement of the shower Monte Carlo simulation program
is also highly desirable.

In practice, the currently used programs\cite{mar2},
based on MLLA, seem to be very successful in describing the hadronic
events in the e$^+$e$^-$-annihilation at current energies.
{}From the theoretical point of view, however,
they are correct only at the next-to-leading order.
For example, the process-independent QCD mass scale
$\Lambda_{\overline{MS}}$ is not used for the effective coupling,
because it requires the accuracy of next-to-next order.
It should also be noted that some of the shortcomings
at the parton level may be compensated
by the phenomenological adjustments in the hadronisation model.
Such adjustments give flexibility to the simulation,
and may give an illusion of having universally successful model.
This freedom, in fact, gives the ambiguities in the analysis,
which would limit the usefulness of the simulation when it is
applied at higher energies.

It is also significant to find the cutoff scale $Q_0$ below which
the perturbative process terminates.
It determines the size of the non-perturbative corrections.

As is discussed above, there are a number of interesting quantities
to be analysed at TRISTAN.
The careful experimental analysis, together with further theoretical
clarification, will make the description of the multiple hadroproduction
in the perturbative QCD more promising.

\end{document}